\def \abc#1#2#3#4 {\reference#1, {\sl#2}, {\bf#3}, #4}
\def \blank {\lower 5pt\hbox to 0.75in{\hrulefill}}

\def \cm{~\rm{cm}}
\def \s{~\rm{s}}
\def \km{~\rm{km}}

\def \g{~\rm{g}}
\def \AU{~\rm{AU}}
\def \yrs{~\rm{yrs}}
\def \yr{~\rm{yr}}

\def \lae{\mathrel{<\kern-1.0em\lower0.9ex\hbox{$\sim$}}}
\def \gae{\mathrel{>\kern-1.0em\lower0.9ex\hbox{$\sim$}}}

\documentclass[12pt,preprint]{aastex}
\begin{document}

\title{ON THE FORMATION OF MULTIPLE-SHELLS \\
     AROUND ASYMPTOTIC GIANT BRANCH STARS}

\shorttitle{Arcs in AGB}
\shortauthors{Soker}

\author{Noam Soker\altaffilmark{1}}
\affil{University of Virginia, Department of Astronomy, P.O.~Box 3818, 
       Charlottesville, VA 22903-0818, USA}
\email{soker@physics.technion.ac.il}

\altaffiltext{1}{On Sabbatical from the University of Haifa at Oranim,
       Department of Physics, Oranim, Tivon 36006, Israel.} 

\begin{abstract}

Two types of models for the formation of semi-periodic concentric multiple
shells (M-shells) around asymptotic giant branch (AGB) stars and in
planetary nebulae are compared against observations. 
Models that attribute the M-shells to processes in an extended
wind acceleration zone around AGB stars result in an optically 
thick acceleration zone, which reduces the acceleration efficiency 
in outer parts of the extended acceleration zone. 
This makes such models an unlikely explanation for the formation of M-shells. 
Models which attribute the M-shell to semi-periodic variation in 
one or more stellar properties are most compatible with observations. 
The only stellar variation models on time scales of $\sim 50-1500 \yrs$     
that have been suggested are based on an assumed solar-like magnetic cycle. 
Although ad-hoc, the magnetic cycle assumption fits naturally
into the increasingly popular view that magnetic activity plays a role in 
shaping the wind from upper AGB stars.
  
\end{abstract}

\keywords{circumstellar matter $-$ planetary nebulae: general $-$ stars
AGB and post-AGB $-$ stars: mass loss}

\clearpage 

\section{INTRODUCTION}
 The arcs and rings which appear in the images of several 
planetary nebulae (PNe) and proto-PNe
(Sahai {\it et al.} 1998b; Kwok, Su, \& Hrivnak 1998;
Su {\it et al.} 1998; Sahai {\it et al.} 1999; Bond 2000;
Hrivnak, Kwok, \& Su 2001, hereafter HKS), 
as well as in one AGB star (IRC+10216; Mauron \& Huggins 1999, 2000),
are thought to be concentric (more or less) semi-periodic shells;
some shells are complete while other are not.
Reviews of the arc and ring properties are given by HKS and 
Kwok, Su \& Stoesz (2001; hereafter KSS).  
 Not all observed arcs and rings are truly concentric, but in all cases 
their ``center'' is relatively (to their radius) close to the central star
(KSS).
 I will use the fact that not all arcs are concentric to constrain 
models for their formation. 
 Like Mauron \& Huggins (2000) and to save space, I term these concentric 
semi-periodic shell multiple-shells, or M-shells. 
The time intervals between consecutive shells change from system to system
and in some cases between shells in the same system, with typical time 
intervals of $t_s \sim 100-1000 \yrs$. 

 Three classes of models were proposed for M-shells formation. 
$(i)$ Binary models (e.g., Harpaz, Rappaport, \& Soker 1997;
Mastrodemos \& Morris 1999) where  the orbital
period sets the time interval between the M-shells. 
 The advantage of this kind of model is that the (semi)-periods are exactly
in the orbital period range where the binary interaction, being tidal 
interaction (Harpaz {\it et al.} 1997) or accretion (Soker 2001), 
turns off and on along an eccentric orbit. 
 However, there are strong arguments against binary models 
(e.g., Soker 2000; Mauron \& Huggins 2000). 
Hence, binary model will not be addressed in this paper, but I do 
allow for the possibility that a companion plays a passive role,
e.g., spining-up the AGB star to initiate magnetic activity. 
$(ii)$ Processes in the circumstellar matter, e.g., instabilities in 
dust-gas coupling (Deguchi 1997; Simis, Icke, \& Dominik 2001). 
The relevant processes in this type of models occur in regions of size 
$r_a$ which are much larger than the stellar radius, $r_a \gg R_\ast$. 
 The basic advantage is that the dynamical time $\sim r_a^{3/2}/(G M)^{1/2}$,
where $M \sim 1 M_\odot$ is the stellar mass, can be set to be of the order
of the intershell time interval if $r_a \sim 20-100 \AU$. 
 In the next section I argue that despite this advantage, there are 
problems with this class of models. 
$(iii)$ Semi-periodic variations in one or more of the stellar
properties, leading to semi-periodic variation in the mass loss rate.
 Helium-shell flashes are ruled out (Sahai {\it et al.} 1998b; 
Kwok {\it et al.} 1998) because their typical inter-flash 
period is $\gtrsim 10^4 \yrs$, while pulsation periods are too short. 
 This led me (Soker 2000) to suggest that a solar-like magnetic activity 
cycle in the progenitor AGB star is behind the formation of M-shells
(see also Garcia-Segura, Lopez, \& Franco 2001). 

 In the present paper I compare models of the above types ($ii$) and ($iii$)
against properties of M-shells, and show that circumstellar 
models suffer from severe problems.
 The stellar semi-periodic variation can account for the M-shell properties,
in particular if a binary companion is present. 
     
\section{CONSTRAINING AND SUPPORTING DIFFERENT MODELS}
\subsection{Radiation Pressure and Opacity}
 In this subsection I show that any model for the M-shell formation which
is based on processes in the circumstellar matter is in conflict  
with wind acceleration by radiation pressure. 
 The relevant process which forms the M-shells must occur before the wind 
reaches its high-Mach terminal speed.
 For simplicity I take a constant acceleration $a$ over a radius 
$r_a$ much larger than the stellar radius $r_a \gg R_\ast$. 
 To form variation over a time interval $t_s$, the acceleration time
of each mass element should be of that order.
 The terminal wind velocity is $v_t = a t_s$, and the wind velocity
in the acceleration zone $R_\ast < r \leq r_a$ is 
$v=(2 a r)^{1/2}=(2 v_t r/t_s)^{1/2}$. 
Assuming spherical symmetry, the wind density is  
\begin{equation}
\rho= \frac {\dot M}{4 \pi r^2 v} = 
\frac {\dot M t_s^{1/2}}{ 4 \pi (2 v_t)^{1/2}} \frac{1}{r^{5/2}},  
    \qquad R_\ast < r \leq r_a. 
\end{equation}
 Since the material in the acceleration zone expands slowly, it is
influenced by the gravity of the central star. 
The radiation pressure must overcome the gravity, since in the wind zone
the pressure gradient is negligible. 
 The total gravitational force on the material in the acceleration zone is
\begin{equation}
F_g = \int_{R_\ast}^{r_a} \frac{GM}{r^2} \rho 4 \pi r^2 dr
= \frac{GM\dot M}{3 R_\ast^{3/2}} \left( \frac {2 t_s}{v_t} \right)^{1/2},
\end{equation}
where equation (1) for $\rho$ was used in the integration.
 Radiation can exert a force $F_{\rm rad} = \beta L/c$, where $c$ is the
speed of light, $L$ the stellar luminosity and $\beta$ is 
the effective average number of times a photon is scattered. 
 Substituting typical values, e.g., for IRC+10216 
(Winters, Dominik, \& Sedlmayr 1994;  Mauron \& Huggins 2000)
we find the constraints on the average number of times a photon is 
scattered in order to support the extended acceleration zone, 
\begin{eqnarray} 
\beta_a = 2.6  
\left( \frac {L} {2 \times 10^4 L_\odot} \right) ^{-1} 
\left( \frac {t_s} {500 \yr } \right)^{1/2} 
\left( \frac {R_\ast} {6 \AU} \right) ^{-3/2} 
\left( \frac {v_t} {15 \km \s^{-1}} \right) ^{-1/2} 
\nonumber \\ \times
\left( \frac {\dot M} {2 \times 10^{-5} M_\odot \yr ^{-1}} \right)  
\left( \frac {M} {0.7 M_\odot} \right) .  
\end{eqnarray} 
 In addition to supplying the counter force in the acceleration zone, 
the radiation supplies the momentum to the wind $\dot M v_t$. 
The average number of scattering required here is 
\begin{equation}
\beta_w = 0.74  
\left( \frac {L} {2\times 10^4 L_\odot} \right) ^{-1} 
\left( \frac {\dot M} {2 \times 10^{-5} M_\odot \yr ^{-1}} \right)  
\left( \frac {v_t} {15 \km \s^{-1}}\right).  
\end{equation}
The total effective average number of times a photon is 
scattered is 
$\beta_t = \beta_a + \beta_w$. 
 For these parameters $\beta_t \simeq 3$, which requires the
wind to be optically thick. 
 Indeed, the optical depth from a radius $R_d< r < r_a$ to infinity,
where $R_d \simeq 10 \AU$ is the dust formation radius, is given by 
\begin{eqnarray} 
\tau = \int _r^{\infty} \kappa \rho dr^\prime = 26 
\left( \frac {t_s} {500 \yr } \right)^{1/2} 
\left( \frac {v_t} {15 \km \s^{-1}} \right) ^{-1/2} 
\left( \frac {\dot M} {2 \times 10^{-5} M_\odot \yr ^{-1}} \right)  
\nonumber \\ \times
\left( \frac {\kappa} {10 \cm^2 \g^{-1}} \right)   
\left( \frac {r} {10 \AU} \right) ^{-3/2} , 
\end{eqnarray}
where $\kappa$ is the dust opacity scaled with a typical value for AGB stars
(Jura 1986; Winters {\it et al.} 2000), and the density was taken from
equation (1). 
 Only at $r \sim 100 \AU$ does the wind become optically thin.
  
 Simis {\it et al.} (2001) take $L=2.4 \times 10^4 L_\odot$ and obtained 
average mass loss rate and velocity of $\dot M \sim 10^{-4} M_\odot \yr^{-1}$
and $v_t \simeq 40 \km \s^{-1}$, respectively, which require $\beta_w= 8$.
This is a very high efficiency for wind acceleration, not found usually 
in AGB stars (Knapp 1986). 
 Such a large average number of times a photon is scattered requires that 
radiative transfer be followed through the optically thick acceleration zone.   
 Simis {\it et al.} (2001) don't follow the radiative transfer, but rather
assume optically thin envelope, making their results questionable.   
(From the density profile in the model of Simis {\it et al.} 
I find that there is a mass of $\sim 0.2 M_\odot$ in the region between 
the stellar surface, $R_\ast=6 \AU$, and the dust 
formation radius, $R_d =15 \AU$. 
This is more than the envelope mass in their stellar model. 
It is not clear what is the influence of this unreasonable assumption on
their results.)

 To summarize, any model based on an extended acceleration zone requires
high efficiency of radiation acceleration extending along the entire
acceleration zone. 
 However, for typical parameters, most, or even all, of the acceleration 
zone is optically thick, because the velocity is low and density high.
 Since radiation reaches outer region only after it is scattered to
longer wavelength, acceleration tends to be less efficient in
outer regions. 
 I therefore consider unlikely that such models can explain the 
formation of M-shells.  
In any case, radiation transfer must be considered in building
such models, 
{{{ as well as the stability of the process 
(Garcia-Segura {\it et al.} 2001).}}} 

\subsection{Non-concentric Arcs}
 In three objects the M-shells are not concentric and/or intersecting 
each other (HKS): AFGL 2688 (the Egg Nebula), NGC 7027, and IRC+10216. 
 Interestingly, the aspherical dense nebula in the center of each of 
these objects departs from axisymmetry:
  In the Egg Nebula the eastern bright region $\sim 6^{\prime \prime}$
from the center observed in the NICMOS image is more extended than the
western bright region on the other side of the central star
(Sahai {\it et al.} 1998a).
 The HST images of NGC 7027 (Latter {\it et al.} 2000) show a more
extended and brighter region to the southwest side along the
equatorial plane. 
 The carbon star IRC+10216 contains an inner elliptical structure on a 
scale of $\sim 10^{\prime \prime}$ (Kastner \& Weinntraub 1994). 
 The variation in the spacing between the contours and polarization
vectors (fig. 2 of Kastner \& Waintraub 1994) are similar in nature
to the variation between the M-shell spacing. 
The departure from axisymmetry appears also at $\sim 0.1^{\prime \prime}$
from the center (Weigelt {\it et al.} 1998; Haniff \& Buscher 1998). 

 A plausible common explanation for both the non-concentric M-shells and
the departure from axisymmetry of the inner bright region is a binary
model (Soker \& Rappaport 2001).     
 The intersecting shells require the orbital period to be shorter, but not
much shorter, than the intershell time interval. 
Each shell then, is ejected while the mass losing AGB star is at a different 
orbital phase, hence moving in a different direction. 
 Since in all three objects the M-shells intersect close to the star,
the orbital velocity of the AGB star around the center of mass
$v_1$ can't be too small (see Soker \& Rappaport 2001).
 If two consecutive shells are ejected when the AGB star moves in opposite
directions, i.e., half an orbital period apart, then the shell will collide 
when at a distance of $r_i \sim \Delta R v_t/(2 v_1)$ from the central star.
 In other cases they will collide further away.
 For these objects I find $r_i \lesssim 5-10 \Delta R$, hence
 $v_1 \gtrsim 0.1-0.05 v_t$ ($v_t$ is the terminal expansion velocity 
of the M-shells). 
 I take as typical values $v_t=15 \km \s^{-1}$, and AGB and companion 
stellar masses of $M_1=0.8 M_\odot$, and $M_2=0.6 M_\odot$  respectively. 
 The constraints on the orbital separation and period are 
(eqs. 1-2 of Soker \& Rappaport 2001) $a \lesssim  100 \AU$ and 
$P_{\rm orb} \lesssim 10^3 \yr$, respectively. 
 This binary model naturally explains the intersecting and/or non-concentric
M-shells in models where the M-shells are formed during a short interval
close to the AGB star.  

 This binary model can't account for the non-concentric M-shells in extended
acceleration zone models, where the formation time of each M-shell is about 
equal to the intershell time interval. 
 The reason is that the required binary orbital period is of the order 
or shorter than the intershell time intervals.
 Hence, either each shell will be smeared, or else each shell have the average
velocity, which is the binary center of mass velocity, and the shells will
not intersect and not be displaced, i.e., they will be concentric. 
 Therefore, another explanation should be found for the presence of 
non-concentric and/or intersecting M-shells in models based on 
an extended acceleration zone.  

 I therefore predict that these three objects contain a binary companion
of mass $\sim 0.5 M_\odot$, either a white dwarf or a main sequence star,
with orbital periods inside the range of $\sim 50-500 \yrs$.  

\subsection{Polar-Disccontinueity}

In some systems there is a clear difference between the M-shell structure
along the symmetry axis and that in the equatorial plane. 
This is usually attributed to different illuminating radiation, e.g., 
the searchlight beams in the Egg Nebula. 
By carefully examining the structure of some M-shells, I find the 
situation to be more complicated. 
($i$) The different illuminating radiation can explain the brighter 
M-shell segments along the polar directions in the Egg Nebula. 
However, the M-shells in the Egg Nebula split, intersect, and possess 
irregular spacing along the polar direction much more so than closer 
to the equator. This is more prominent along the northern search light beams. 
 This effect can't be attributed to different illumination.  
($ii$) In IRAS 20028+3910 (HKS) there are three short arcs on each side
along the symmetry axis, with the inner one being at $\sim 1 ^{\prime \prime}$.
However, there is faint nebulosity in other directions from the central 
star of IRAS 20028+3910, up to $\sim 1.5 ^{\prime \prime}$ from the central 
star in the SW direction (Ueta, Meixner, \& Bobrowsky 2000), where no arcs
are observed (HKS). 
 I therefore argue that the presence of arcs along the symmetry axis but  
not, or much less prominent, in other direction in IRAS 20028+3910, is real. 

 The physical differences between equatorial and polar segments
of M-shell that I argue for here are very tentative. 
 In any case, it is interesting that magnetic activity, i.e., surface spots, 
are known to have polar and/or equatorial concentration, as well as 
uniform distribution (e.g., Schrijver \& Title 2001 and references therefin).
 Magnetic activity can naturally explain polar to equatorial variation
{{{ (Gracia-Segura {\it et al.} 2011)}}},
if they turn out to be real. 

\section{SUMMARY}

 The goal of this paper is to examine two types of models for the formation 
of semi-periodic concentric multiple shells (termed here M-shells;
also called arcs) against observations. 
The comparison of models with observations suffers from the small 
number of objects with well defined M-shells.  
Binary models were not discussed here, as they were found in previous works
to suffer from severe problems (Soker 2000; Mauron \& Huggins 2000).
In particular, the inter-shell time intervals are not constant. 
The binary model can be `saved' only if another non-periodic component 
is added to the periodic orbital motion. 
I know of no mechanism to cause such a non-periodic variation component
which can act together with the orbital period  
(beside the magnetic cycle which can act by itself; see below). 
 
In the present paper I find that models that attribute the M-shells to
processes in an extended, up to $\sim 100 \AU$, wind acceleration zone,
result in an optically thick acceleration zone. 
This reduces the acceleration efficiency in the outer parts of the
extended acceleration zone, making such models unlikely. 
Simis {\it et al.} (2001) for example, assume that the acceleration
zone is optically thin; hence, I consider their results questionable. 

 Models which attribute the M-shell to semi-periodic variation in 
one or more stellar properties are most compatible with observations. 
The departure of the M-shells from being concentric in some systems
can be easily explained by an orbital motion of the AGB star around
the center of mass with a stellar companion.
The stellar companion does not play a dynamical role in the formation 
process itself, but it may spin-up the AGB star to stimulate an efficient 
dynamo. 
 The only stellar variation models on time scales of $\sim 50-1500 \yrs$     
that have been suggested are based on a magnetic cycle 
(Soker 2000; Garcia-Segura 2001).
From main sequence stars it is well known that in addition to a, more or 
less, uniform surface activity, magnetic fields, e.g., spots, can have
polar and/or equatorial concentrations. 
 This may explain the tentative claim I made here that in some
objects there are structural differences between M-shell segments 
along the polar directions and equatorial plane. 
Although the magnetic cycle is an ad-hoc assumption, it fits naturally
into the increasingly popular view that magnetic activity plays a role in 
shaping the wind from upper AGB stars (e.g., 
{{{ Chevalier \& Luo 1994; Rozyczka \& Franco 1996;}}} Soker 2000; 
Gardiner \& Frank 2001).
  
\acknowledgements
I thank Robert Link for comments on the original manuscript.
This research was supported in part by grants from the 
US-Israel Binational Science Foundation.

\end{document}